\documentclass[aps,prl,twocolumn]{revtex4} 
\usepackage{amsmath,amsfonts,amssymb}
\usepackage{graphicx,epic,eepic}
\begin{document}
\title{\bf Massive phonon modes from a BEC-based analog model}
\author{Matt Visser}
\email{matt.visser@mcs.vuw.ac.nz}
\author{Silke Weinfurtner}
\email{silke.weinfurtner@mcs.vuw.ac.nz}
\affiliation{School of Mathematics, Statistics, and Computer Science,
Victoria University of Wellington,
PO Box 600, Wellington, New Zealand.}

\begin{abstract}
  
  Two-component BECs subject to laser-induced coupling exhibit a
  complicated spectrum of excitations, which can be viewed as two
  interacting phonon modes. We study the conditions required to make
  these two phonon modes decouple. Once decoupled, the phonons not
  only can be arranged travel at different speeds, but one of the
  modes can be given a mass --- it exhibits the dispersion relation of
  a massive relativistic particle $\omega = \sqrt{\omega_0^2 + c^2
    k^2}$.  This is a new and unexpected excitation mode for the
  coupled BEC system. Apart from its intrinsic interest to the BEC
  community, this observation is also of interest for the ``analogue
  gravity'' programme, as it opens the possibility for using BECs to
  simulate massive relativistic particles in an effective ``acoustic
  geometry''.

\bigskip

\centerline{cond-mat/0409639}

\bigskip

\centerline{24 September 2004; \LaTeX-ed \today }

\end{abstract}
\maketitle

\enlargethispage{250pt}
\clearpage
\newcommand{\norm}[1]{\left\Vert#1\right\Vert}
\newcommand{\betrag}[1]{\left\vert#1\right\vert}
\newcommand{\bra}[1]{\langle #1\vert}
\newcommand{\ket}[1]{\arrowvert #1 \rangle}
\newcommand{\braket}[1]{\langle #1\rangle}
\newcommand{\kin}{-\frac{\hbar^2}{2m} \, \nabla^2 \,}
\newcommand{\oppdag}[2]{\hat{#1}^\dagger(t,\vec{#2})}
\newcommand{\opp}[2]{\hat{#1}(t,\vec{#2})}
\newcommand{\bog}{\hat{\Psi}(t,\vec{x})=\Phi(t,\vec{x}) \, + \, \varepsilon \hat{\psi}(t,\vec{x})\, + \, \ldots}
\newcommand{\zeit}{i \,\hbar \,  \frac{\partial}{\partial t} \,}
\newcommand{\mad}{\sqrt{\rho(t,\vec{x})}e^{i\theta(t,\vec{x})}e^{-\frac{i\mu t}{\hbar}}}

Analogue models for gravitation can be used to simulate classical and
quantum field theory in curved space-time.  The first analogue model
for black holes, and for simulating Hawking evaporation was suggested
by Bill Unruh \cite{Unruh}.  He demonstrated that a sound wave
propagating though a converging fluid flow exhibits the same
kinematics as light does in the presence of a curved space-time
background.  Since then several other media have been analyzed, and
the field has developed tremendously. The first approach specifically
using Bose--Einstein condensates as an analogue model was made
nineteen years later \cite{Garay}.  Since then various configurations
of BECs have been studied to simulate different scenarios for
gravity~\cite{BLV-BEC,Fedichev,BLV-probe,silke-arbeit,silke2,Uwe:new}.
Until now it has only been possible to simulate light, (generally
speaking, massless relativistic particles), propagating through a
curved space-time~\cite{Ergo,normal,normal2,book,vortex}. In the
following a two-species BEC is used to extend the class of equations
that can be simulated to the full curved-space Klein--Gordon
equations.  In the language of the BEC community --- we have developed
a way of giving a mass to the phonon.


The system we will use in our theoretical analysis is an ultra cold
2-component BEC atomic gas. For example, a two-component condensate of
${}^{87}$Rb atoms in different hyperfine levels, which we label
$\ket{A}$ and $\ket{B}$.  (Experiments using two different spin
states, $\ket{F=1,m=-1}$ and $\ket{F=2,m=2}$, were first performed at
\textit{JILA} in 1999~\cite{Myatt:1997}.)  At zero temperature nearly
all atoms occupy the ground state. Then the quantized field describing
the microscopic system can be replaced by a classical mean-field, a
macroscopic wave-function.  In this so-called mean-field approximation
the number of non-condensed atoms is small. Interactions between the
condensed and non-condensed atoms are neglected in the mathematical
description, but two-particle collisions between condensed atoms are
included.  In the case of a two-component system, interactions within
each species ($U_{AA}$, $U_{BB}$) and between the different species
($U_{AB}=U_{BA}$) take place.  In addition the two condensates are
coupled by a laser-beam, which drives transitions between the two
hyperfine states with a constant rate $\lambda$.  (Without the
coupling $\lambda$ no mass term is generated, which is consistent
with~\cite{Uwe:new}.)  The resulting coupled time-dependent
Gross--Pitaevskii equations are:
\begin{widetext}
\begin{equation}
 \label{2GPE}
\begin{split}
 i \, \hbar \, \partial_{t} \Psi_{A} =& \left[ -\frac{\hbar^2}{2\,m_{A}} \nabla^2 + V_{A}^* + U_{AA} \, \betrag{\Psi_{A}}^2  + U_{AB} \betrag{\Psi_{B}}^2 \right]  \Psi_{A} + \lambda \, \Psi_{B} \, ,\\
 i \, \hbar \, \partial_{t} \Psi_{B} =& \left[ -\frac{\hbar^2}{2\,m_{B}} \nabla^2 + V_{B}^* + U_{BB} \, \betrag{\Psi_{B}}^2  + U_{AB} \betrag{\Psi_{A}}^2 \right] \Psi_{B} +\,\lambda \, \Psi_{A} \, ,
\end{split}
\end{equation}
\end{widetext}
where $V_{A/B}^*=V_{A/B}-\mu_{A/B}$ denotes the combined effects of
the external potential $V_{A/B}$ and the chemical potential
$\mu_{A/B}$~\cite{Garcia-Ripoll:dec2003,Jenkins-Kennedy:nov2003}.  In
the eikonal approximation the two stationary background states are
described by their densities $\{\rho_{A},\rho_{B}\}$ and phases
$\{\theta_{A},\theta_{B}\}$:
\begin{equation}
\Psi_{X}= \sqrt{\rho_{X} }\; e^{i(\theta_{X})} \quad\hbox{for}\quad X=A,B \, .
\end{equation}
These four variables are in general not  independent of time and space.

In the following, we study zero sound in the overlap region of the
two-component system, produced by exciting density perturbations which
are small compared to the density of each condensate cloud.  In the
first experiment studying localized excitations in a one-component
Bose--Einstein condensate \cite{Ketterle:28july1997}, a laser beam was
used to generate a small modulation in the density. Using a
phase-contrast imaging it was shown that the resulting perturbation
corresponds to a sound wave. The observed speed of sound is
\begin{equation} \label{c}
c(r)=\sqrt{\frac{4 \pi \, \hbar^2 \, a \, \rho(r) }{m^2}}=\sqrt{\frac{U \, \rho(r)}{m} }\, ,
\end{equation} 
where $\rho(r)$ is the density of the ground state, $a$ is the
scattering length, $m$ is the atomic mass and $U$ is the
self-interaction constant.  The mathematical equations describing
these perturbation leads to the well-known hydrodynamic equations,
which are the basis for the most fruitful of the analogies between
condensed matter physics and general relativity~\cite{Unruh,Garay,Ergo,book}.

The same method can be used to obtain the kinematic equations for
small perturbations propagating in a two-component system.  Given that
the density modulation is small, the perturbations in the densities
and phases can be linearized around their unperturbed macroscopic
states $\{\rho_{A0},\rho_{B0}\}$:
\begin{equation}
\Psi_{X}= \sqrt{\rho_{X0}+ \varepsilon \, \rho_{X1} }\, e^{i(\theta_{X0}+  \varepsilon \, \theta_{X1} )} 
\quad\hbox{for}\quad X=A,B \, .
\end{equation}
These states still satisfy the coupled Gross--Pitaevskii equation.
After a straightforward calculation the terms of first order in
$\varepsilon$ representing the sound waves include two coupled
equations for the perturbation of the phases
\begin{equation} \label{soundwave1}
\begin{split}
\dot{\theta}_{A1}=
&-\vec{v}_{A0} \nabla \theta_{A1}  - \frac{\tilde{U}_{AA}}{\hbar} \, \rho_{A1} +  \frac{\tilde{U}_{AB}}{\hbar} \, \rho_{B1}, \\
\dot{\theta}_{B1}=
&-\vec{v}_{B0} \nabla \theta_{B1}  - \frac{\tilde{U}_{BB}}{\hbar} \, \rho_{B1} +  \frac{\tilde{U}_{AB}}{\hbar} \, \rho_{A1}.  \\
\end{split} 
\end{equation}
Here $\vec{v}_{A0}=({\hbar}/{m_{A}}) \nabla \theta_{A0}$ and
$\vec{v}_{B0}=({\hbar}/{m_{B}}) \nabla \theta_{B0}$ are the background
velocities for the macroscopic ground states, while
\begin{eqnarray}
\tilde{U}_{AA} &=&   U_{AA} -\frac{\lambda}{2}  \frac{\sqrt{\rho_{B0}}}{\sqrt{\rho_{A0}}^{3}} ,
\nonumber
\\
\tilde{U}_{BB} &=&   U_{BB} -\frac{\lambda}{2}  \frac{\sqrt{\rho_{A0}}}{\sqrt{\rho_{B0}}^{3}} ,
\\
\tilde{U}_{AB} &=& U_{AB}- \frac{\lambda}{2}  \; \frac{1}{\sqrt{\rho_{A0} \, \rho_{B0} }},
\nonumber
\end{eqnarray}
are modified interaction potentials for the two coupled condensates.
In addition to these two phase equations, there are two coupled
equations for the density perturbations
\begin{widetext}
\begin{equation} \label{soundwave2}
\begin{split}
\dot{\rho}_{A1}=&
- \nabla\left(\frac{\hbar}{m_{A}} \,\rho_{A0} \nabla \theta_{A1} + \rho_{A1}   \vec{v}_{A0}\right) 
+ \frac{2\lambda}{\hbar}\sqrt{\rho_{A0} \, \rho_{B0}} \; (\theta_{B1}-\theta_{A1}), \\
\dot{\rho}_{B1}=&
- \nabla\left(\frac{\hbar}{m_{B}} \, \rho_{B0} \nabla \theta_{B1} + \rho_{B1}   \vec{v}_{B0}\right) 
+ \frac{2\lambda}{\hbar}\sqrt{\rho_{A0}\,\rho_{B0}} \; (\theta_{A1}-\theta_{B1}) .\\
\end{split} 
\end{equation}
\end{widetext}
It is useful to define the coupling matrix
\begin{equation}
\Xi=\frac{1}{\hbar}
\left(
\begin{array}{rr}
\tilde{U}_{AA} & - \tilde{U}_{AB}  \\
-\tilde{U}_{AB} & \tilde{U}_{BB}  \\
\end{array}
\right) \, .
\end{equation} 
A second coupling matrix can be introduced as
\begin{equation}
\Lambda=\frac{2\sqrt{\rho_{A0}\,\rho_{B0}} }{\hbar}
\left(
\begin{array}{rr}
+\lambda & -\lambda \\ -\lambda & +\lambda \\
\end{array}
\right) \, ,
\end{equation}
which vanishes completely if the coupling laser is switched off.  Last
but not least, it is also useful to introduce the background velocity
matrix $V$ (a $2\times2$ matrix of 3-vectors)
\begin{equation}
V=
\left(
\begin{array}{rr}
\vec{v}_{A0}  & 0 \\ 0 &  \vec{v}_{B0} \\
\end{array}
\right),
\end{equation}
and the mass-density matrix $D$
\begin{equation}
D=\hbar
\left(
\begin{array}{rr}
\frac{\rho_{A0}}{m_{A}}  & 0 \\ 0 &  \frac{\rho_{B0}}{m_{B}}   \\
\end{array}
\right) .
\end{equation}

Collecting terms into a $2\times2$ matrix equation, the equations for
the phases (\ref{soundwave1}) and densities (\ref{soundwave2}) become
\begin{equation} \label{thetavecdot}
\dot{\bar{\theta}}=  -\,\Xi \cdot \bar{\rho} - V \cdot \nabla \bar{\theta},
\end{equation} 
\begin{equation} \label{rhovecdot}
\dot{\bar{\rho}}= \, - \nabla \, \left( D \cdot \nabla \bar{\theta} + V \cdot  \bar{\rho}   \right) 
- \Lambda \cdot \bar{\theta} \, ,
\end{equation} 
where ${\bar{\theta}}^{\,T} = (\theta_{A1},\theta_{B1})$ and 
${\bar{\rho}}^{\,T} = (\rho_{A1},\rho_{B1})$.\\

Equation (\ref{thetavecdot})  can be used to
eliminate $\bar{\rho}$ and $\dot{\bar{\rho}}$ in equation (\ref{rhovecdot}), leaving us with a single  matrix equation for
the perturbed phases:
\begin{eqnarray} \label{phaseequation}
 &\partial_{t} \left(\Xi^{-1} \cdot \dot{\bar{\theta}} \, \right) =
 - \partial_{t} \left(\Xi^{-1} \cdot V \cdot \nabla \bar{\theta} \, \right) 
 - \nabla       \left(V \cdot \Xi^{-1} \cdot \dot{\bar{\theta}} \, \right)  
 \nonumber
 \\
&\qquad 
 + \nabla \left[ \left(D - V \cdot \Xi^{-1} \cdot V \right) \nabla \bar{\theta} \, \right] 
 + \Lambda \cdot \bar{\theta}  
\end{eqnarray}
This equation tells us how a localized collective excitation in a
two-component system with a permanent coupling --- if $\lambda$ is
nonzero --- develops in time. So far there are no restrictions on the
masses $(m_{A},m_{B})$, densities $(\rho_{A0},\rho_{B0})$, background
velocities $(\vec{v}_{A0},\vec{v}_{B0})$, interaction constants
$(U_{AA},U_{BB},U_{AB})$, and coupling constant $(\lambda)$.  In the
eikonal approximation this differential equation leads to the Fresnel
equation
\begin{eqnarray}
\label{fresnel}
&&\det\big\{
\omega^2 \; \Xi^{-1}  -  \omega \left[\Xi^{-1} \cdot (V \mathbf{k}) +  (V \mathbf{k}) \cdot \Xi^{-1} \right]
\nonumber
\\
&&
\qquad
- [ D \; k^2 - (V \mathbf{k}) \cdot \Xi^{-1} \cdot  (V \mathbf{k}) ] + \Lambda ]
\big\}
=0\,,
\end{eqnarray}
which is in general a quartic dispersion relation for two interacting
phonon modes.  Here $  V \mathbf{k}$ denotes the matrix
\begin{equation}
 V \mathbf{k} =
\left(
\begin{array}{cc}
\vec{v}_{A0}  \cdot \vec k & 0 \\ 0 &  \vec{v}_{B0} \cdot \vec k \\
\end{array}
\right).
\end{equation}

The first step in analyzing equation (\ref{phaseequation}) is to ask
whether it is possible to decouple the system into two independent
phonon modes. We have found decoupling is not possible without
introducing several constraints on the background quantities.

Focusing on the last term in equation (\ref{phaseequation}), the
eigenvectors for a non-zero coupling $\lambda \neq 0$ are given by
$\{[1,1] ,[-1,1]\}$ and the corresponding eigenvalues are $\{0,{4 \,
  \lambda \sqrt{\rho_{A0}\rho_{B0}}}/{\hbar}\}$. The eigenvectors are
fixed and independent of any of the other physical variables.  As a
result the only way to decouple equation (\ref{phaseequation}) into
two phonon modes, is to decompose it in the following way:
\begin{equation}  \label{eigenstates}
\bar{\theta}= 
\left(\begin{array}{r} \tilde{\theta}_{1} \\ \tilde{\theta}_{1} \end{array} \right)
+
\left(\begin{array}{r} -\tilde{\theta}_{2} \\ \tilde{\theta}_{2} \end{array} \right)
\end{equation} 
We now analyze equation (\ref{phaseequation}) term by term with
respect to this decomposition.

The term on the LHS has the same eigenvectors as equation
(\ref{eigenstates}) if and only if $\tilde{U}_{AA}=\tilde{U}_{BB}$.
The eigenvalues of $\Xi^{-1}$ corresponding to $\{[1,1] ,[-1,1]\}$ are
$\{ \hbar / (\tilde{U}_{AA}-\tilde{U}_{AB}),\hbar /
(\tilde{U}_{AA}+\tilde{U}_{AB}) \}$.  This places another constraint
on the interaction variables: $\tilde{U}_{AA} \neq \tilde{U}_{AB}$.
(Further discussion regarding this tight requirement will be presented
later.)

The eigenvectors for the first two terms on the RHS in equation
(\ref{phaseequation}) can be found in one step, by using the fact that
they have simultaneous eigenvectors if and only if the commutator
$[\Xi^{-1},V]=0$ vanishes.  Therefore
$\vec{v}_{A0}=\vec{v}_{B0}=\vec{v}_{0}$, the background velocities
must be equal if the two phonon modes are to decouple.  The
corresponding eigenvalues are those from the matrix $\Xi^{-1}$
multiplied by $\vec{v}_{0}$. In other words, the backgrounds of two
condensates must be in phase $\theta_{A0}=\theta_{B0}$.

We are now left with the penultimate term in equation
(\ref{phaseequation}).  Because $V=\vec{v}_{0} \cdot \mathbf{1} $, and
the eigenvectors of $\Xi^{-1}$ are already known, there is only the
mass-density matrix $D$ to consider. The last constraint to decouple
the equation for the two phases then is $\hbar \,
\rho_{A0}/m_{A}=\hbar \, \rho_{B0}/m_{B}=d$. So $D=d\cdot \mathbf{1}$
has for both eigenvectors the eigenvalue $d = \hbar \sqrt{\rho_{A0} \;
  \rho_{B0}}/\sqrt{m_A\;m_B}$.

Applying all this to equation (\ref{phaseequation}) one obtains two
decoupled equations for the phonon modes described by the eigenstates
(\ref{eigenstates}):
\begin{equation}
\partial_\mu (f^{\mu\nu}_I \;\partial_\nu \tilde\theta_I) = 
{4\lambda\sqrt{\rho_{A0}\rho_{B0}}\over \hbar} \; \delta_{2I}\; \tilde\theta_I,
\quad\hbox{for}\quad
I=1,2.
\end{equation}
Here
\begin{equation}
f^{\mu \nu}_I={d\over c^2_I}
\left(
\begin{array}{c|c}
-1         & -v_0^j \\
\hline
-v_0^j & c_I^2\;  \delta^{ij} - v_0^i \; v_0^j
\end{array}\right) \, ,
\end{equation}
where the propagation speeds are defined in terms of the eigenvalues $\Xi_I$ of the matrix $\Xi$ by
\begin{equation} \label{cis}
c_{I}^2=   \Xi_I \; d =  \frac{d \, (\tilde{U}_{AA} + (-1)^{I} \tilde{U}_{AB})}{\hbar} \, .
\end{equation}

Introducing effective ``spacetime metrics'' by the identifications
$\sqrt{-g_I} \; g_I^{\mu\nu} = f_I^{\mu\nu}$ and $g_I = 1/\det[
g_I^{\mu\nu}]$, we can recast these equations as a pair of
curved-space Klein-Gordon [massive d'Alembertian] equations
\begin{equation} \label{Klein-Gordon equation}
\frac{1}{\sqrt{-g_I}}
\partial_{\mu} \left( \sqrt{-g_I} \, g^{\mu \nu}_{I} \partial_{\nu} \, \tilde{\theta}_{I} \right) 
- \mathbf{\mathring{m}^2 } \,  \delta_{2\, I} \,  \tilde{\theta}_{I} =0.
\end{equation} 
Here
\begin{equation}
g^{\mu \nu}_I=  \left({d\over c_I}\right)^{-2/(D-2)} 
\left[  {1\over c^2_I} 
\left(
\begin{array}{c|c}
-1         & -v_0^j \\
\hline
-v_0^j & c_I^2 \delta^{ij} - v_0^i \; v_0^j
\end{array}\right) \right],
\end{equation}
which depends on the space-time dimension $D$~\cite{Ergo,vortex} in
such a manner that
\begin{equation}
g_{\mu \nu}^I=  \left({d\over c_I}\right)^{2/(D-2)} \; 
\left(
\begin{array}{c|c}
-(c_I^2 - v_0^2)         & -v_0^j \\
\hline
-v_0^j & \delta^{ij}
\end{array}\right)\, .
\end{equation}
Finally the mass-term is
\begin{equation}
\mathbf{\mathring{m}^2 }= 
- \frac{4 \, \lambda \sqrt{\rho_{A0} \, \rho_{B0}}}{\hbar} 
\; \left({  c_2^2/d^D  }\right)^{1/(D-2)} ,
\end{equation}
corresponding to the natural oscillation frequency
\begin{eqnarray}
\omega_0^2 
&=& 
\mathbf{\mathring{m}^2 } \; c_2^2 \left({d\over c_2}\right)^{2/(D-2)}
= 
-{4\lambda\sqrt{\rho_{A0}\rho_{B0}} \; c_2^2 \over \hbar\; d} 
\nonumber
\\
&=&
-{4\lambda \sqrt{m_A \; m_B} \; c_2^2\over \hbar^2}.
\end{eqnarray}
The dispersion relation (in the eikonal limit) is then
\begin{equation}
(\omega - \mathbf{v_0 }\cdot \mathbf{k})^2 - c_I^2 \; k^2 = \omega_0^2\; \delta_{2I}\,.
\end{equation}
(A similar calculation, but restricted to a one-condensate system,
where all variables are allowed to be time and space dependent, but no
mass term is present, has been presented in \cite{silke-arbeit}.)
Equation (\ref{Klein-Gordon equation}) is the curved-space
Klein--Gordon equation (massive d'Alembertian equation).  For
$\tilde{\theta}_{1}$, (corresponding to perturbations in the two
condensates $A$ and $B$ oscillating ``in phase''), the mass term is
always zero. However, for a laser-coupled system ($\lambda\neq0$) the
mass-term in the equation for $\tilde{\theta}_{2}$, (corresponding to
perturbations in the two condensates $A$ and $B$ oscillating ``in
anti-phase''), does not vanish.

Comparing the definition for the speed of sound (\ref{c}) in a one
component system, with the $c_I$ introduced here, we see that the
$c_{I}$ (\ref{cis}) are the modified speeds of sound for each phonon
mode. (If $U_{AB}=\lambda=0$ the two condensates decouple and the
$c_I$ limit to the independent phonon speeds in each condensate
cloud.)  This fact leaves us with the possibility of constructing two
different types of analog model. So far we have been dealing with a
two-metric structure, which is interesting in itself~\cite{normal2,vsl}. For instance, in
the absence of laser-coupling ($\lambda=0$) the presence of two
different speed of sounds can be used for tuning
effects~\cite{Uwe:new}.

If we wish to more accurately simulate the curved spacetime of our own
universe, another constraint should be placed on the system, to make
the two speeds of sound equal $c_{1}=c_{2}$.  This yields a single
sound-cone structure, to match the observed fact that our universe
exhibits a single light-cone structure.  This condition is fulfilled
if we set $\tilde{U}_{AB}=0$.  While the in-phase perturbations will propagate
exactly at the speed of sound, 
\begin{equation}
\mathbf{v_{s}} =\mathbf{v_0} + \mathbf{\hat k}\;c_I \, ,
\end{equation}
the anti-phase perturbations will move
with a lower group velocity given by:
\begin{equation}
\mathbf{v_{g}}= \frac{\partial \omega}{\partial \mathbf{k}}
=\mathbf{v_0} + \mathbf{\hat k}\;\frac{c_I^2}{\sqrt{\omega_0^2\; \delta_{2I} + c_I^2\; k^2}}
\, .
\end{equation}
Here $\mathbf{k}$ is the usual wave number. This explicitly
demonstrates that the group velocity of the anti-phase eigenstate
depends on the laser-induced coupling between the condensates.


In conclusion, the calculation presented in this article is of
interest to two separate communities. For the BEC community, it
provides a specific example of how to tune an interacting 2-BEC
condensate in such a manner as to obtain a massive phonon. Without the
fine tuning, it provides an example of two interacting phonon modes
whose dispersion relation is governed by the quartic Fresnel equation
(\ref{fresnel}). For the general relativity community, this article
provides an example of an analogue system that can be used to mimic a
minimally coupled scalar field embedded in a curved spacetime.


\emph{Acknowledgements:} This research was supported by the Marsden
Fund administered by the Royal Society of New Zealand. We wish to thank Crispin Gardiner, Piyush Jain, and Ashton Bradley  for their thoughtful comments.



\end{document}